\documentclass{appolb}
\usepackage{epsfig}

\begin{document}
\title{Measurement of D$^0$ v$_2$ in Pb-Pb collisions at $\sqrt{s_{NN}}=2.76$~TeV with ALICE at the LHC%
\thanks{Presented at Strangeness in Quark Matter 2011, Cracow}%
}
\author{Chiara Bianchin for the ALICE Collaboration
\address{University \& INFN - Padova, Italy}
}
\maketitle

\begin{abstract}
The study particle azimuthal anisotropy in heavy-ion collisions provides insight on the collective hydrodynamic expansion of the system and on its equation of state. The measurement of the elliptic flow, $v_2$, of D mesons compared to that of light hadrons is expected to be sensitive to the degree of thermalization of charm quarks within the quark-gluon plasma.
The first measurement of D$^0$ meson $v_2$ with the ALICE detector at the LHC will be presented. The preliminary results obtained with the first Pb-Pb run at LHC show a hint of non-zero $v_2$ in $2<p_{\rm t} <5 $ GeV/$c$.
\end{abstract}
  
\section{Introduction}
The ALICE experiment \cite{aliceJINST} at the LHC is a dedicated heavy ion experiment with the goal of studying the quark-gluon plasma (QGP).

In  heavy-ion collisions with non-zero impact parameter the fireball exhibits an azimuthal asymmetry with respect to the reaction plane, defined by the azimuth of the impact parameter and the beam direction. Due to collective effects, this initial geometrical anisotropy evolves with time into an asymmetry in momentum space, due to different pressure gradients in-plane and out-of-plane. The final particle azimuthal distribution, therefore, is influenced by the initial geometry. It can be evaluated via the 2nd coefficient of the Fourier expansion of the azimuthal distribution, called elliptic flow ($v_2$) \cite{lhcpred}. The measurement of the heavy flavour particle $v_2$ is sensitive in particular to the degree of thermalization of the system, probing the interaction of heavy quarks, produced in the early stages of the nuclear collision, with the QGP. The heavy-flavour hadronization mechanism is also investigated \cite{hadmec,molnar}.

Some models \cite{molnar} predict charm elliptic flow to be smaller than that of light quarks at $p_{\rm t}$ up to $\sim$2 GeV/$c$ and comparable at higher $p_{\rm t}$. So far, the heavy-quark flow has been measured  via non-photonic electrons by the PHENIX experiment \cite{phenix}. ALICE can contribute with a direct measurement through the fully reconstructed hadronic decays of D mesons, thanks to its vertex reconstruction, tracking and particle identification capabilities.

The measurement of the elliptic flow is complementary to the measurement of the nuclear modification factor ($R_{AA}$). ALICE measured \cite{raaD} a suppression in the D meson yield in Pb-Pb collisions with respect to pp, rescaled by the number of binary collisions,  which probes a strong interaction of charm quarks with the QGP via energy loss mechanism. Although the systematics uncertainties are large, the D mesons suppression is at the same level as that of light charged hadrons.

\section{The first measurement of D$^0\,v_2$}
This analysis is based on  about $17\times10^6$ minimum bias events collected during the LHC 2010 Pb-Pb run. The trigger was defined as the logical OR of the VZERO, which consist of two scintillator arrays located at the two sides of the interaction point and on the Silicon Pixel Detector (SPD), which is the innermost detector in the central rapidity region.
The centrality is determined via a Glauber-model fit of the VZERO signal amplitude \cite{plamen, albericaQM}. The centrality class used for this measurement is between 30 and 50\% of the Pb-Pb nuclear cross section.

The first D$^0 \, v_2$ measurement was performed with event plane methods and 2-particle cumulants \cite{cumulants}, due to limited statistics.

The analysis strategy consists in the selection of the D$^0 \rightarrow \rm K^- \pi^+$ candidates (described in \S \ref{subsec:sigextr}) followed by the $v_2$ measurement with the methods mentioned above (\S \ref{subsec:evpl}, \ref{subsec:qc}). The estimation of the systematic uncertainties is described in \S \ref{subsec:syst} and finally in \S \ref{sec:results} the results are shown.

\subsection{Signal extraction}\label{subsec:sigextr}
A typical signature of the D$^0\rightarrow \rm K^- \pi^+$ decay is the presence of two tracks with opposite charges from a secondary vertex pointing the primary vertex and with an invariant mass around the D$^0$ meson mass (1.865 GeV). The necessary spatial resolution for the determination  of the primary and secondary vertices is assured by the Inner Tracking System (ITS) and the excellent track resolution is provided by the Time Projection Chamber (TPC) and the ITS.
For each pair of tracks a secondary vertex is defined as their point of closest approach. The significance of the signal is optimized by selecting track pairs ({\it candidates}) that pass specific cuts on the decay length, the angle between the D$^0$ meson flight line and its reconstructed momentum, and the decay track impact parameter \cite{analD}.
The identification of the kaon rejects additional background, mostly at low $p_{\rm t}$. The identification relies on the specific energy deposit dE/dx in the TPC and on the time of flight of the particle measured by the Time Of Flight detector (TOF).

\subsection{Event plane methods}\label{subsec:evpl}
The event plane methods correlate each candidate's azimuthal angle ($\phi$) with the reaction plane angle ($\Psi_{RP}$). The reaction plane is estimated through the event plane $\Psi_n$ determined by the so-called Q-vector which is a weighted sum of the azimuthal angles of all the TPC tracks in $|\eta|<0.8$ with quality requirements including 70 clusters in the TPC and a $\chi^2/n.d.f<4$. To remove autocorrelation, the tracks coming from the $\rm D^0$ candidate under study are subtracted from the Q-vector. The $\Psi_2$ is used for this analysis:  
\begin{equation}
 Q_2={\sum_{i=0}^{N} w_i \cos 2\phi_i \choose \sum_{i=0}^{N} w_i \sin 2\phi_i} \qquad \Psi_2 = \frac{1}{2} \tan^{-1} \left(\frac{Q_{2,y}}{Q_{2,x}}\right).
\end{equation}
In our case, the $w_i$ are $\phi$-weights determined on a run-by-run basis.
The precision on the measured event plane is limited by the finite number of reconstructed tracks. The resolution on $\Psi_2$ was estimated with the two sub-events method \cite{eventplane} and the resulting correction factor $\sigma_{\Psi_2}$ ($\left\langle \cos2(\Psi_2 - \Psi_{RP}) \right\rangle$), to be applied to the observed $v_2^{obs}$ ($v_2=v_2^{obs}/\sigma_{\Psi_2}$), is 0.91.

The three methods used to extract the ${\rm D}^0 \,\,v_2$ in 3 bins of $p_{\rm t}$ are described in the following sub-sections.
\subsubsection{D$^0$ yield in two bins of azimuthal angle}\label{ssub:massfit}
The first method consists in comparing the signal yield in bins of $\Delta\phi = \phi-\Psi_2$. The signal is extracted with an invariant mass analysis. Two ranges of $\Delta\phi$ are defined: $ \left[-\pi/4, \pi/4 \right) \bigcup \left[3\pi/4, 5\pi/4 \right) $, which gives the in-plane yield ($N_{IN}$) and $\left[\pi/4, 3\pi/4 \right) \bigcup \left[5\pi/4, 7\pi/4  \right)$, which gives the out-of-plane yield ($N_{OUT}$). The yield is extracted from the subtraction of the background below the exponential fit function from the total fit, in a region of $3\sigma$ around the mean of the Gaussian fit (Fig.~\ref{fig:2Dmeth1}, left panel).
Integrating the Fourier expansion in the two $\Delta\phi$ ranges, $v_2^{obs}$ can be expressed as:
\begin{equation}\label{f:v2inout}
 v_2^{obs}=\frac{\pi}{4} \frac{N_{IN}-N_{OUT}}{N_{IN}+N_{OUT}}
\end{equation}

\subsubsection{Method of the side band subtraction}
\begin{figure}
\centering
  \includegraphics[width=0.9\textwidth]{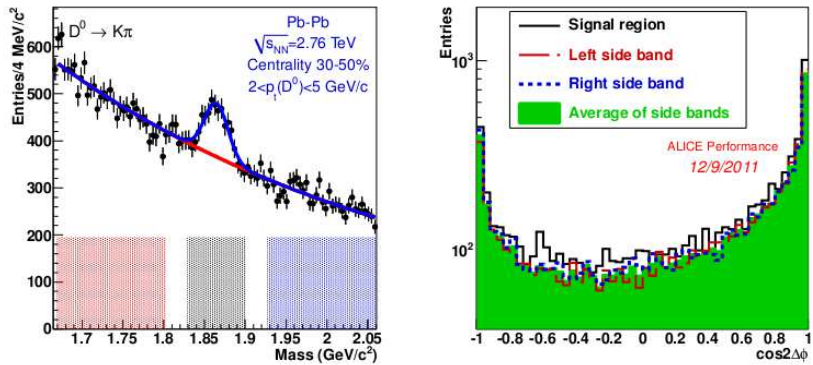}

\caption{Left panel: Invariant mass spectrum with shaded bands defining the side-bands (red and blue) and the peak region (black) and fit functions. Right panel: Distribution of $\cos2\Delta\phi$ in different mass regions. The green area is the average of the left- and right-hand side-bands. $p_{\rm t}$ bin $2<p_{\rm t}<5~$GeV$/c$.}\label{fig:2Dmeth1}
 \end{figure}

The $\cos2\Delta\phi$ distribution of D$^0$ candidates in three mass ranges is considered: two side bands, which are then averaged, and the peak region,  as shown in shaded areas in Fig.~\ref{fig:2Dmeth1}, left panel. The distribution of the averaged side bands, shown by the green filled graph in Fig.~\ref{fig:2Dmeth1}, right panel, normalized to the background fitting function under the D$^0$ peak, is subtracted from the $\cos2\Delta\phi$ in the peak region (solid black line) and finally the $v_2^{obs}$ is computed as $\left\langle \cos2\Delta\phi \right\rangle$.

\subsubsection{Fit of $v_2$ vs mass}\label{ssub:fitvsmass}
 Using the information of the fit to the invariant mass spectrum (Gaussian plus exponential), the fraction of signal (S) and background (B) over the total is estimated as a function of mass. The dependence of $v_2^{obs}=\left\langle \cos2\Delta\phi \right\rangle$ measured in small bins of mass (M) is drawn in Fig.~\ref{fig:v2cum}, left panel, for $2<p_{\rm t}<5$ GeV/$c$. The fitting function is:
\begin{equation}\label{f:fitfunc}
 \frac{B(M)}{S(M)+B(M)} \cdot v_2^{bkg}(M) + \frac{S(M)}{S(M)+B(M)} \cdot v_2^{sig}
\end{equation}
where $v_2^{bkg}(M)=p_0 M+p_1$ and $v_2^{sig}=p_2$ are obtained from the fit free parameters $p_0$, $p_1$, $p_2$.

\subsection{Cumulants}\label{subsec:qc}
\begin{figure}
 \includegraphics[width=0.5\textwidth,height=0.23\textheight]{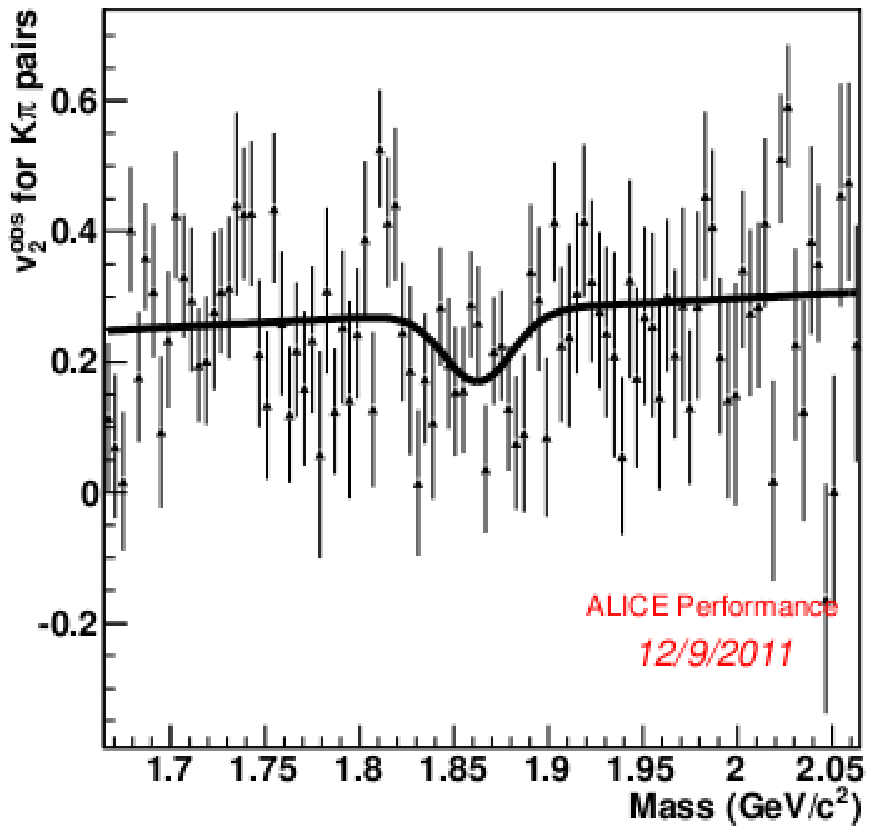}
 \includegraphics[width=0.5\textwidth]{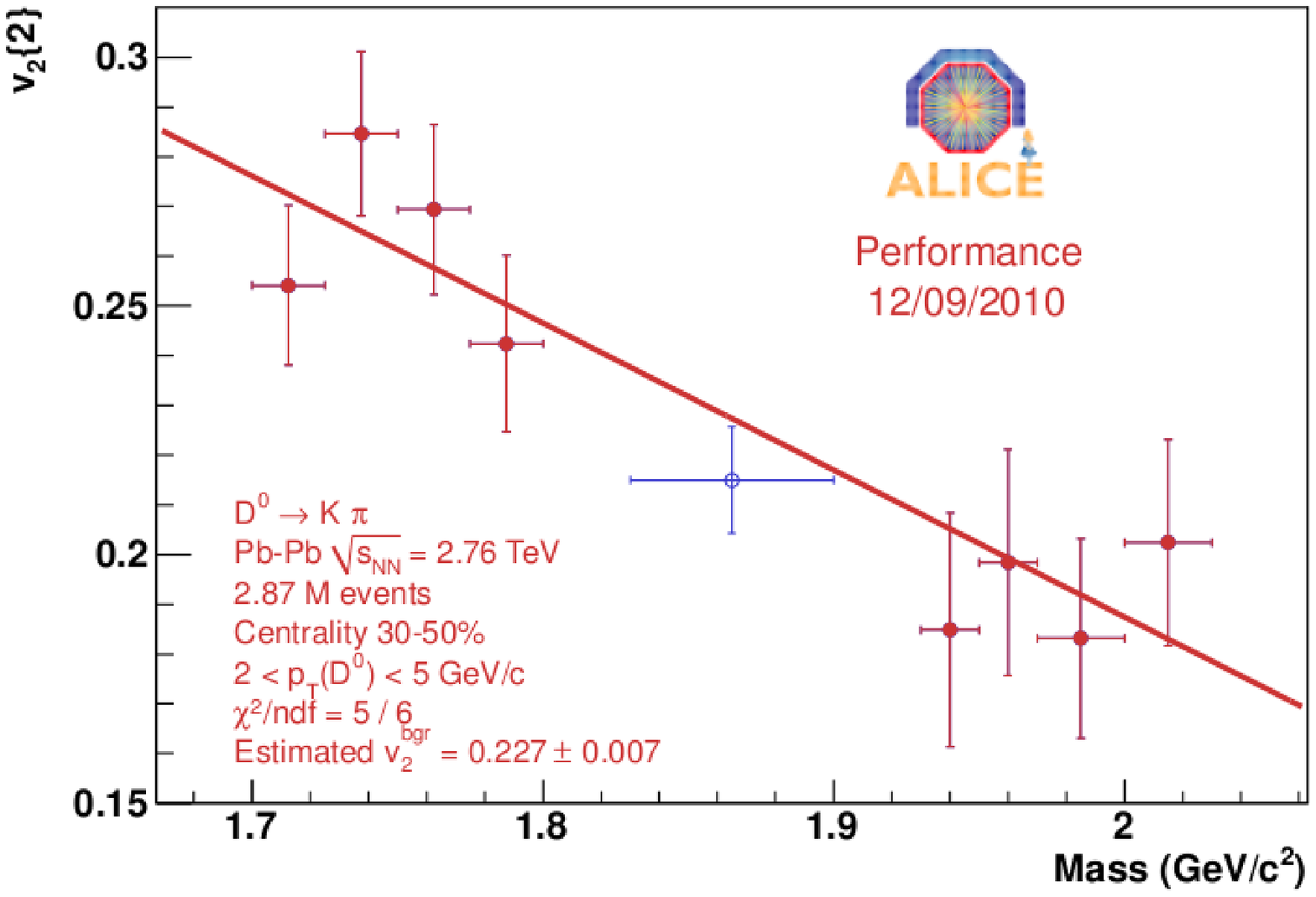}
\caption{Left panel: $v_2^{obs}$ vs K$\pi$ invariant mass fitted with Eq.~\ref{f:fitfunc} (\S \ref{ssub:fitvsmass}). Right panel: Elliptic flow measured with 2-particle cumulants as a function of mass. The red closed circles correspond to the side bands mass bins and the blue open circle to the peak region}\label{fig:v2cum}
\end{figure}

A 2-particle cumulants, the Q-cumulants method \cite{qcum} was used. It was not possible to use 4-particle cumulants due to limited statistics. The 2-particle cumulants are calculated as ${\rm QC\{2\}}= \left\langle \rm e^{2 i(\phi_1 - \phi_2)} \right\rangle$ and are related to the elliptic flow by $v_2\{2\}={\rm QC\{2\}}$. In Fig.~\ref{fig:v2cum}, right panel, showing $v_2\{2\}$ vs mass for $2<p_{\rm t}<5$~GeV/$c$, the closed markers are interpolated with a line and the background flow extracted ($v_2^{bkg}$) is subtracted from the measured $v_2$ in the mass peak region ($v_2^{peak}$) as follows:
\begin{equation}
 v_2^{sgn}=\frac{S(M)+B(M)}{S(M)}v_2^{peak}- \frac{B(M)}{S(M)}v_2^{bkg}
\end{equation}
where S, B are respectively the signal and background obtained from the fit to the invariant mass distribution in 3$\sigma$ around the measured D$^0$ mass.

\subsection{Systematic uncertainties}\label{subsec:syst}
A 2.5\% systematic uncertainty  due to the different event plane resolution within the centrality class is accounted for all event plane methods.\\
For the method of the yield extraction in two $\Delta\phi$ bins a $\sim 20\%$ uncertainty is induced by the signal extraction procedure. A systematic ranging from 20\% up to a factor 2-3 is estimated from the cut variation. The method was validated using charged tracks and comparing to official ALICE results.\\
The systematic uncertainty for the method that uses side band subtraction, estimated by varying the width of the side bands and their binning ranges from 20 to 50\%, depending on $p_{\rm t}$. In the $v_2$ vs mass method, the fraction of signal and background as well as the mass bin width were varied, giving a systematic uncertainty of about 30-40\% $p_{\rm t}$-wide.\\ 
The QC\{2\} method is sensitive to the variation of the background fit function and to the D$^0$ selection cuts which give a total systematic uncertainty from 40\% to 100\% from low to high $p_{\rm t} $.

\section{Results}\label{sec:results}
\begin{figure}
 \includegraphics[width=0.5\textwidth]{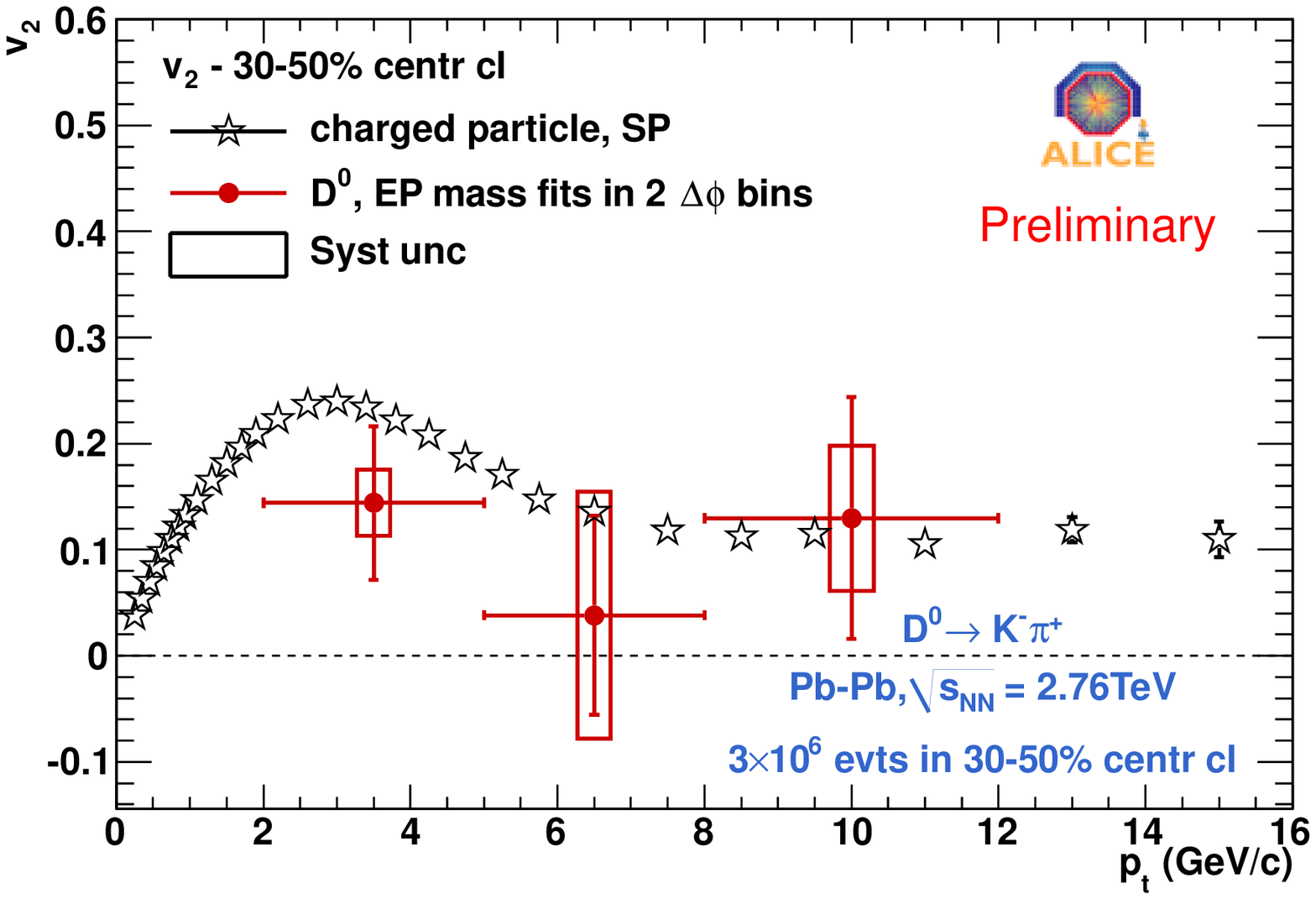}
 \includegraphics[width=0.5\textwidth]{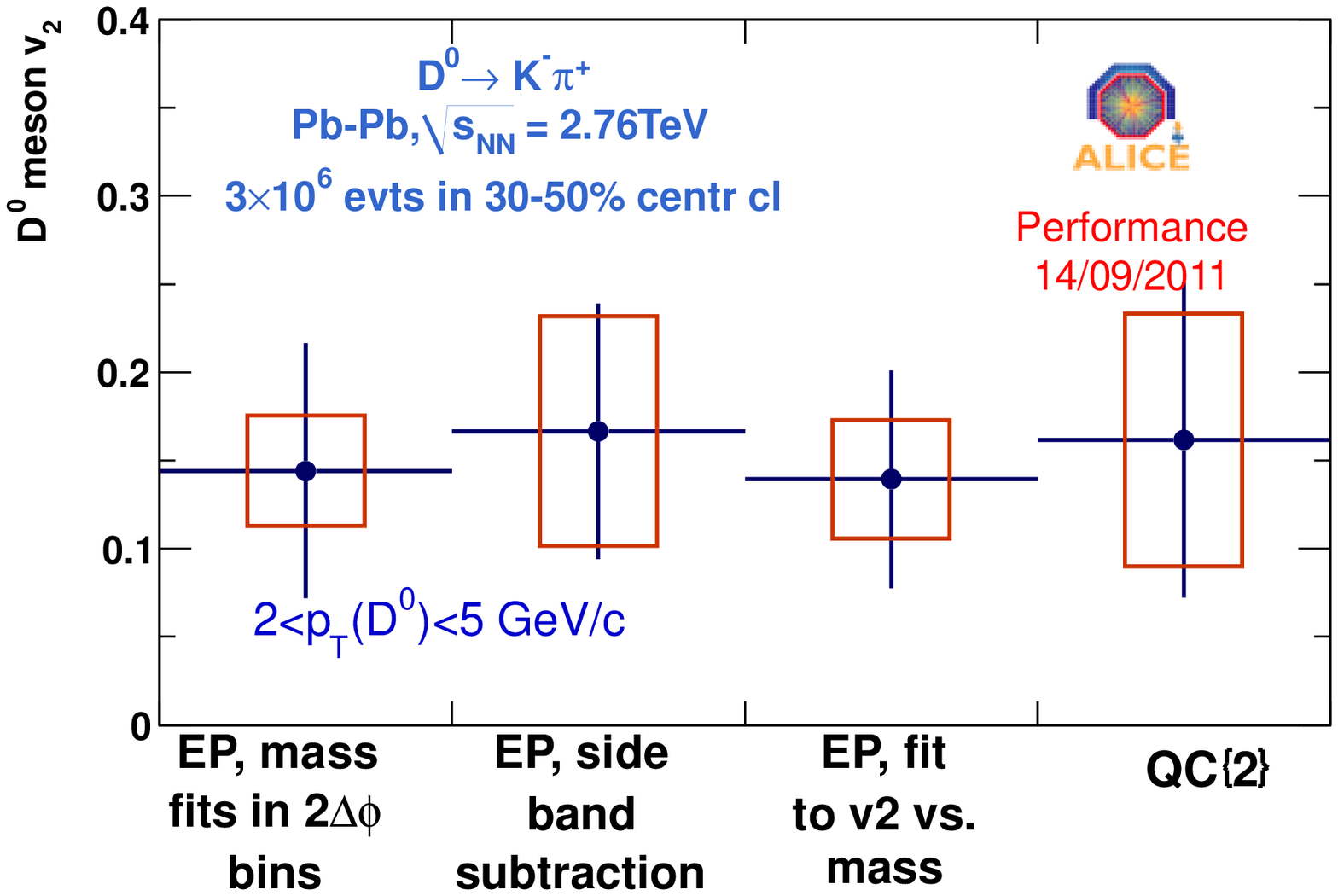}
\caption{Left panel: $v_2$ vs $p_{\rm t}$ with the 2$\Delta\phi$ bins method (\S \ref{ssub:massfit}), boxes are systematic uncertainties. Right panel: $v_2$ in $2<p_{\rm t}<5~$GeV$/c$ with all methods described.}\label{fig:results}
\end{figure}

In Fig.~\ref{fig:results}, left panel, the D$^0$ $v_2$ measured with the event plane method of the two $\Delta\phi$ bins (full circle) for three $p_{\rm t}$ intervals and the $v_2$ for charged tracks measured by ALICE (empty star) are shown. The right panel shows the $v_2$ from all methods in the first $p_{\rm t}$ bin, where errors are the smallest, and they are consistent. The measurement is performed for inclusive D$^0$, no feed-down subtraction was considered so far.

Within 1.8$\sigma$ a non-zero $v_2$ is measured for D$^0$ mesons in the $p_{\rm t}$ range 2-3 GeV/$c$ for Pb-Pb collisions at 2.6 TeV. 

\end{document}